# The total irregularity of a graph

July 1, 2018

**Hosam Abdo, Darko Dimitrov**

*Institut für Informatik, Freie Universität Berlin,*
*Takustraße 9, D–14195 Berlin, Germany*
E-mail: [abdo,darko]@mi.fu-berlin.de

**Abstract**

In this note a new measure of irregularity of a simple undirected graph $G$ is introduced. It is named the *total irregularity* of a graph and is defined as $\mathrm{irr}_t(G) = \frac{1}{2} \sum_{u,v \in V(G)} |d_G(u) - d_G(v)|$, where $d_G(u)$ denotes the degree of a vertex $u \in V(G)$. The graphs with maximal total irregularity are determined. It is also shown that among all trees of same order the star graph has the maximal total irregularity.

## 1 Introduction

We consider only finite, undirected graphs without loops or multiple edges. For a graph $G$, we denote by $n = |V(G)|$ and $m = |E(G)|$ its order and size, respectively. For $v \in V(G)$, the degree of $v$, denoted by $d_G(v)$, is the number of edges incident to $v$. By $N_G(u)$, we denote the set of vertices that are adjacent to a vertex $u$, and by $\overline{N}_G(u)$ the set of vertices that are not adjacent to $u$. A sequence of non-negative integers $d_1, ..., d_n$ is a *graphic sequence*, or a *degree sequence*, if there exists a graph with the vertex set $\{v_1, ..., v_n\}$ such that $d(v_i) = d_i$. A *pendant* vertex is a vertex of degree one. A *universal* vertex is the vertex adjacent to all other vertices. A set of vertices is said to be *independent* when the vertices are pairwise non-adjacent. The vertices from an independent set are *independent vertices*.

A graph is *regular* if all its vertices have the same degree, otherwise it is *irregular*. However, it is of interest to measure how irregular it is. Several approaches have been proposed that characterize how irregular a graph is.

Albertson [5] defines the *imbalance* of an edge $e = uv \in E(G)$ as $|d_G(u) - d_G(v)|$ and the *irregularity* of $G$ as

$$\mathrm{irr}(G) = \sum_{uv \in E(G)} |d_G(u) - d_G(v)|. \qquad (1)$$

He presented upper bounds on irregularity for bipartite graphs, triangle-free graphs and arbitrary graphs, as well as a sharp upper bound for trees. Some claims about bipartite graphs given in [5] have been formally proved in [16]. Related to the work of Albertson is the work of Hansen and Mélot [15], who characterized the graphs with $n$ vertices and $m$ edges with maximal irregularity. For more results on imbalance, the irregularity of a graph, and other approaches, that characterize how irregular a graph is, we redirect the reader to [3, 4, 7, 8, 9, 10, 11, 12, 14, 17, 18, 19].

In the sequel we introduce and consider an irregularity measure that is related to the irregularity measure (1). As well as (1), the new measure also captures the irregularity only by a single parameter, namely the degree of a vertex, and for a graph $G$ it is defined as

$$\mathrm{irr}_t(G) = \frac{1}{2} \sum_{u,v \in V(G)} |d_G(u) - d_G(v)|. \qquad (2)$$

Because of the obvious connection with the irregularity of a graph, we called it the *total irregularity* of a graph. Note that the total irregularity of a given graph is completely determined by its degree sequence – graphs with the same degree sequences have the same total irregularity, which is an expected property of an irregularity measure. However, this is not always true with the irregularity of a graph (see Figure 1 for such an example).

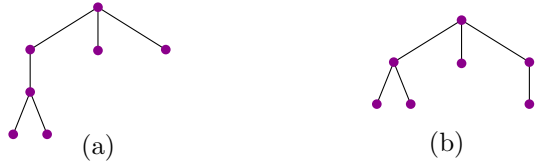

**Figure 1:** Two non-isomorphic graphs $G_1$ and $G_2$ with the same degree sequence $1, 1, 1, 1, 2, 3, 3$. They have different irregularities ($\mathrm{irr}(G_1) = 10$ and $\mathrm{irr}(G_2) = 8$), but the same total irregularity ($\mathrm{irr}_t(G_1) = \mathrm{irr}_t(G_2) = 22$).

Obviously, both measures are zero if and only if $G$ is regular, and $\mathrm{irr}_t(G)$ is an upper bound of $\mathrm{irr}(G)$. Very recently, these two measurements were compared in [1], where it was shown that for a connected graph $G$ with $n$ vertices, $\mathrm{irr}_t(G) \leq n^2 \mathrm{irr}(G)/4$. Moreover, if $G$ is a tree, then it was shown that $\mathrm{irr}_t(G) \leq (n-2)\mathrm{irr}(G)$. In this note, we focus on graphs with maximal total irregularity.

## 2 Graphs with maximal total irregularity

Let $G_{max}$ be a graph with $n$ vertices and with maximal $\mathrm{irr}_t$. Assume that $G_{max}$ has $q$ universal vertices, where $0 \leq q < n$ (the case $q = n$ is excluded because then $\mathrm{irr}_t(G) = 0$). We denote by $U$ the set of universal vertices of $G_{max}$. Let $\overline{U} = \{\bar{u}_1, \bar{u}_2, \ldots, \bar{u}_{n-q}\}$ the set of non-universal vertices of $G_{max}$. We assume that $d(\bar{u}_1) \geq d(\bar{u}_2) \geq \cdots \geq d(\bar{u}_{n-q-1}) \geq d(\bar{u}_{n-q})$.

**Proposition 2.1.** *Let $\bar{u}_i, \bar{u}_j \in \overline{U}$, $i < j$. Then,*

(a) *there is an edge between $\bar{u}_i$ and $\bar{u}_j$, if $i + j < n - 2q + 1$;*

(b) there is no edge between $\bar{u}_i$ and $\bar{u}_j$, if $i + j > n - 2q + 1$;

(c) inserting or deleting an edge $\bar{u}_i\bar{u}_j$ from $G_{max}$ does not change $\text{irr}_t(G_{max})$, if $i + j = n - 2q + 1$.

*Proof.* (a) Assume that $G_{max}$ does not contain an edge $\bar{u}_i\bar{u}_j$, where $i + j < n + 1 - 2q$. We add such an edge $\bar{u}_i\bar{u}_j$, obtaining a graph $G_a$. The degrees of both vertices $\bar{u}_i$ and $\bar{u}_j$ increase by one. The change of the total irregularity between $\bar{u}_i$ and the universal vertices is $\sum_{x \in U} |d_{G_a}(x) - d_{G_a}(\bar{u}_i)| - \sum_{x \in U} |d_{G_{max}}(x) - d_{G_{max}}(\bar{u}_i)| = -q$, and the change of the total irregularity between $\bar{u}_j$ and the universal vertices is $\sum_{x \in U} |d_{G_a}(x) - d_{G_a}(\bar{u}_j)| - \sum_{x \in U} |d_{G_{max}}(x) - d_{G_{max}}(\bar{u}_j)| = -q$. For the change of the total irregularity between $\bar{u}_i$ and the vertices in $\overline{U}$, it holds that $\sum_{\bar{u}_k \in \overline{U}, k<i} |d_{G_a}(\bar{u}_k) - d_{G_a}(\bar{u}_i)| - \sum_{\bar{u}_k \in \overline{U}, k<i} |d_{G_{max}}(\bar{u}_k) - d_{G_{max}}(\bar{u}_i)| \geq -i + 1$, and $\sum_{\bar{u}_k \in \overline{U}, k>i} |d_{G_a}(\bar{u}_k) - d_{G_a}(\bar{u}_i)| - \sum_{\bar{u}_k \in \overline{U}, k>i} |d_{G_{max}}(\bar{u}_k) - d_{G_{max}}(\bar{u}_i)| \geq n - q - i - 1$. Similarly, for the change of the total irregularity between $\bar{u}_j$ and the vertices in $\overline{U}$, it holds that $\sum_{\bar{u}_k \in \overline{U}, k<j} |d_{G_a}(\bar{u}_k) - d_{G_a}(\bar{u}_j)| - \sum_{\bar{u}_k \in \overline{U}, k<j} |d_{G_{max}}(\bar{u}_k) - d_{G_{max}}(\bar{u}_j)| \geq -j + 2$, and $\sum_{\bar{u}_k \in \overline{U}, k>j} |d_{G_a}(\bar{u}_k) - d_{G_a}(\bar{u}_j)| - \sum_{\bar{u}_k \in \overline{U}, k>j} |d_{G_{max}}(\bar{u}_k) - d_{G_{max}}(\bar{u}_j)| \geq n - q - j$. Thus,

$$\begin{aligned} \text{irr}_t(G_a) &\geq \text{irr}_t(G_{max}) - q - (i-1) + (n - q - i - 1) - q - (j - 2) + (n - q - j) \\ &= \text{irr}_t(G_{max}) + 2(n - 2q + 1 - i - j) \\ &> \text{irr}_t(G_{max}), \end{aligned}$$

which contradicts the assumption that $G_{max}$ is a graph with maximal $\text{irr}_t$.

(b) Assume that $G_{max}$ contains an edge $\bar{u}_i\bar{u}_j$ such that $i + j > n - 2q + 1$. We delete such an edge $\bar{u}_i\bar{u}_j$, obtaining a graph $G_b$. Similarly as in (a), we have

$$\begin{aligned} \text{irr}_t(G_b) &\geq \text{irr}_t(G_{max}) + q + (i-1) - (n - q - i - 1) + q + (j - 2) - (n - q - j) \\ &= \text{irr}_t(G_{max}) + 2(-n + 2q - 1 + i + j) \\ &> \text{irr}_t(G_{max}), \end{aligned}$$

which is a contradiction to the fact that $G_{max}$ is a graph with maximal $\text{irr}_t$.

(c) Assume that $G_{max}$ does not contain an edge $\bar{u}_i\bar{u}_j$ such that $i + j = n - 2q + 1$. We add an edge $\bar{u}_i\bar{u}_j$, where $i + j = n - 2q + 1$, to $G_{max}$, obtaining a graph $G_c$. From (a) and (b), it follows that $d(\bar{u}_k)$ is strictly bigger than $d(\bar{u}_i)$, for all $k < i$. Thus, we have

$$\begin{aligned} \text{irr}_t(G_c) &= \text{irr}_t(G_{max}) - q - (i - 1) + (n - q - i - 1) - q - (j - 2) + (n - q - j) \\ &= \text{irr}_t(G_{max}) + 2(n - 2q + 1 - i - j) \\ &= \text{irr}_t(G_{max}). \end{aligned}$$

□

In the sequel, to simplify the notation we denote $N_{G_{max}}(\bar{u}_1) \cup \{\bar{u}_1\}$ by $N$, and we use $\overline{N}$ instead of $\overline{N}_{G_{max}}(\bar{u}_1)$. By Proposition 2.1, we have that $\bar{u}_1$ is adjacent to all vertices $\bar{u}_i$, $i < n - 2q$, it is not adjacent to all vertices $\bar{u}_i$, $i > n - 2q$, and it might be adjacent to $\bar{u}_{n-2q}$. Therefore, we have the following corollary.

**Corollary 2.1.** $|\overline{N}| = q$ or $|\overline{N}| = q + 1$.

Now, we determine the maximal value of $\text{irr}_t$ of general graphs.

**Theorem 2.1.** *For any simple, undirected graph $G$, $\text{irr}_t(G) \leq \frac{1}{12}(2n^3 - 3n^2 - 2n + 3)$.*

*Proof.* By Proposition 2.1(c), adding or deleting edges $\bar{u}_i\bar{u}_j$, where $i + j = n - 2q + 1$, does not change $\text{irr}_t(G_{max})$. Thus, further we consider that $G_{max}$ does not contain these edges. Then, by Corollary 2.1, it follows that $|\overline{N}| = q + 1$.

The degrees of the vertices in $N$ are as follows: $d(\bar{u}_i) = n - q - 1 - i$, for $i = 1, \ldots, \lceil(n - 2q - 1)/2\rceil$, and $d(\bar{u}_i) = n - q - i$, for $i = \lceil(n - 2q - 1)/2\rceil + 1, \ldots, n - 2q - 1$. All vertices in $\overline{N}$ have degree $q$. The vertices in $U$ are universal and they have degree $n - 1$.

The contribution between the vertices from $U$ and $N$ to $\text{irr}_t(G_{max})$ is

$$\sum_{u_i \in U} \sum_{\bar{u}_j \in N} |d(u_i) - d(\bar{u}_j)|$$

$$= q \left( \sum_{i=1}^{\lceil \frac{n-2q-1}{2} \rceil} (n - 1 - (n - q - 1 - i)) + \sum_{i=\lceil \frac{n-2q-1}{2} \rceil + 1}^{n-2q-1} (n - 1 - (n - q - i)) \right)$$

$$= \frac{1}{2} q \left( (n-2)(n-2q-1) + 2 \left\lceil \frac{n-2q-1}{2} \right\rceil \right). \tag{3}$$

The contribution between the vertices from $U$ and $\overline{N}$ to $\text{irr}_t(G_{max})$ is

$$\sum_{u_i \in U} \sum_{\bar{u}_j \in \overline{N}} |d(u_i) - d(\bar{u}_j)| = q(q+1)(n-1-q). \tag{4}$$

The contribution between the vertices from $N$ and $\overline{N}$ is

$$\sum_{\bar{u}_i \in N} \sum_{\bar{u}_j \in \overline{N}} |d(\bar{u}_i) - d(\bar{u}_j)|$$

$$= (q+1) \left( \sum_{i=1}^{\lceil \frac{n-2q-1}{2} \rceil} (n - q - 1 - i - q) + \sum_{i=\lceil \frac{n-2q-1}{2} \rceil + 1}^{n-2q-1} (n - q - i - q) \right)$$

$$= \frac{1}{2}(q+1) \left( (n-2q)(n-2q-1) - 2 \left\lceil \frac{n-2q-1}{2} \right\rceil \right). \tag{5}$$

Finally, the contribution between the vertices from $N$ to $\text{irr}_t(G_{max})$ is

$$\sum_{\bar{u}_i \in N} \sum_{\bar{u}_j \in N} |d(\bar{u}_i) - d(\bar{u}_j)|$$

$$= \sum_{i=1}^{\lceil \frac{n-2q-1}{2} \rceil - 1} \sum_{j=i+1}^{\lceil \frac{n-2q-1}{2} \rceil} (n - q - i - 1 - (n - q - j - 1))$$

$$+ \sum_{i=1}^{\lceil \frac{n-2q-1}{2} \rceil} \sum_{j=\lceil \frac{n-2q-1}{2} \rceil + 1}^{n-2q-1} (n - q - i - 1 - (n - q - j))$$

$$+ \sum_{i=\lceil \frac{n-2q-1}{2} \rceil+1}^{n-2q-2} \sum_{j=i+1}^{n-2q-1} (n-q-i-(n-q-j))$$

$$= \sum_{i=1}^{\lceil \frac{n-2q-1}{2} \rceil-1} \sum_{j=i+1}^{\lceil \frac{n-2q-1}{2} \rceil} (j-i) + \sum_{i=1}^{\lceil \frac{n-2q-1}{2} \rceil} \sum_{j=\lceil \frac{n-2q-1}{2} \rceil+1}^{n-2q-1} (j-i-1)$$

$$+ \sum_{i=\lceil \frac{n-2q-1}{2} \rceil+1}^{n-2q-2} \sum_{j=i+1}^{n-2q-1} (j-i)$$

$$= \frac{1}{6}(n-2q)(n-2q-1)(n-2q-2) - \left((n-2q-1) - \left\lceil \frac{n-2q-1}{2} \right\rceil\right)\left\lceil \frac{n-2q-1}{2} \right\rceil. \quad (6)$$

After simplifying the sum of (3), (4), (5), and (6), we have

$$\mathrm{irr}_t(G_{max}) = \begin{cases} \frac{1}{12}(2n^3 - 3n^2 - 2n - 4q^3 + 4q) & \text{n even,} \\ \frac{1}{12}(2n^3 - 3n^2 - 2n - 4q^3 + 4q + 3) & \text{n odd.} \end{cases} \quad (7)$$

The maxima of the right side expressions in (7) are obtained for $q = 1$. Thus, finally we have

$$\mathrm{irr}_t(G_{max}) = \begin{cases} \frac{1}{12}(2n^3 - 3n^2 - 2n) & \text{n even,} \\ \frac{1}{12}(2n^3 - 3n^2 - 2n + 3) & \text{n odd.} \end{cases}$$

□

In Figure 2, graphs with maximal total irregularity for $n = 4, 5, 6, 7, 8$ are depicted.

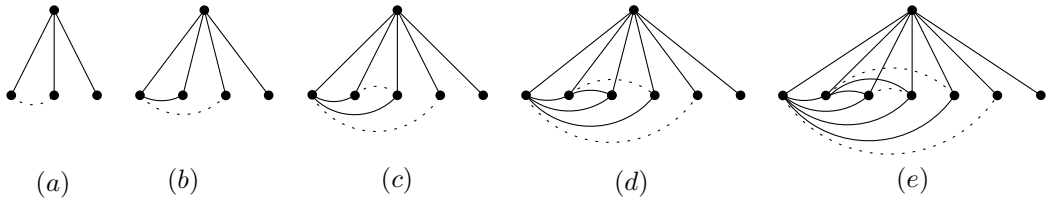

(a)  (b)  (c)  (d)  (e)

**Figure 2:** Graphs with maximal total irregularity with (a) 4, (b) 5, (c) 6, (d) 7, (e) 8 vertices. The dotted edges are optional.

There are $\lfloor \frac{n}{2} \rfloor - 1$ optional edges in $G_{max}$ (edges $\bar{u}_i \bar{u}_j$ that satisfy $i + j = n - 1$ and do not change $\mathrm{irr}_t(G_{max})$). Thus, the number of graphs of order $n$ with the maximal total irregularity is $2^{\lfloor \frac{n}{2} \rfloor - 1}$.

**Proposition 2.2.** *Let $G$ be a tree with n vertices. Then, $\mathrm{irr}_t(G) \leq (n-1)(n-2)$. Moroever, equality holds if and only if $G$ is a star graph.*

*Proof.* Let $G$ be a tree that is not a star, with $u$ as a vertex with maximal degree. Consider a pendant vertex $v$ that is not adjacent to $u$, and is a adjacent to a vertex $w$. We remove the edge $vw$ and add the edge $uv$, obtaining a graph $G'$. After this replacement, only the degrees of $u$ and $w$ change, namely, $d_{G'}(u) = d_G(u) + 1$ and $d_{G'}(w) = d_G(w) - 1$. Thus, we have

$$|d_{G'}(u) - d_{G'}(w)| - |d_G(u) - d_G(w)| = 2,$$

$$\sum_{x \in V(G) \setminus \{u\}} |d_{G'}(w) - d_{G'}(x)| - \sum_{x \in V(G) \setminus \{u\}} |d_G(w) - d_G(x)| \geq -n + 2, \text{ and}$$

$$\sum_{x \in V(G) \setminus \{w\}} |d_{G'}(u) - d_{G'}(x)| - \sum_{x \in V(G) \setminus \{w\}} |d_G(u) - d_G(x)| = n - 1.$$

From the above relations, we obtain $\mathrm{irr}(G') - \mathrm{irr}(G) = 2 - n + 2 + n - 1 = 3$, and therefore $\mathrm{irr}(G') > \mathrm{irr}(G)$. If $G'$ is not the star, then we repeat the above replacement until the resulting graph is the star. The irregularity of the star graph of order $n$ is $(n-1)(n-2)$. $\square$

# References

[1] Darko Dimitrov, Riste Škrekovski, *Comparing the irregularity and the total irregularity of graphs*, submitted.

[2] Yousef Alavi, Jiuqiang Liu, Jianfang Wang, *Highly irregular digraphs*, Discrete Math. **111** (1993) 3–10.

[3] Y. Alavi, A. Boals, G. Chartrand, P. Erdős, O. R. Oellermann, *k-path irregular graphs*, Congr. Numer. **65** (1988) 201–210.

[4] Y. Alavi, G. Chartrand, F. R. K. Chung, P. Erdős, R. L. Graham, O. R. Oellermann, *Highly irregular graphs*, J. Graph Theory **11** (1987) 235–249.

[5] Michael O. Albertson, *The irregularity of a graph*, Ars Comb. **46** (1997) 219–225.

[6] F. K. Bell, *A note on the irregularity of graphs*, Linear Algebra Appl. **161** (1992) 45–54.

[7] F. K. Bell, *On the maximal index of connected graphs*, Linear Algebra Appl. **144** (1991) 135–151.

[8] Y. Caro, R. Yuster, *Graphs with large variance*, Ars Comb. **57** (2000) 151–162.

[9] G. Chartrand, P. Erdős, O. R. Oellermann, *How to define an irregular graph*, Coll. Math. J. **19** (1988) 36–42.

[10] G. Chartrand, K. S. Holbert, O. R. Oellermann, H. C. Swart, *F-Degrees in graphs*, Ars Comb. **24** (1987) 133–148.

[11] L. Collatz, U. Sinogowitz, *Spektren endlicher Graphen*, Abh. Math. Sem. Univ. Hamburg **21** (1957) 63–77.

[12] D. Cvetković, P. Rowlinson, *On connected graphs with maximal index*, Publications de l'Institut Mathematique (Beograd) **44** (1988) 29–34.

[13] G. H. Fath-Tabar, *Old and new Zagreb indices of graphs*, MATCH Commun. Math. Comput. Chem. **65** (2011) 79–84.

[14] D. E. Jackson, R. Entringer, *Totally segregated graphs*, Congress. Numer. **55** (1986) 159–165.

[15] P. Hansen, H. Mélot, *Variable neighborhood search for extremal graphs. 9. Bounding the irregularity of a graph*, DIMACS Ser. Discrete Math. Theoret. Comput. Sci. **69** (2005) 253–264.

[16] Michael A. Henning, Dieter Rautenbach, *On the irregularity of bipartite graphs*, Discrete Math. **307** (2007) 1467–1472.

[17] D. Rautenbach, *Propagation of mean degrees*, Electr. J. Comb. **11** (2004) N11.

[18] D. Rautenbach, L. Volkmann, *How local irregularity gets global in a graph*, J. Graph Theory **41** (2002) 18–23.

[19] D. Rautenbach, I. Schiermeyer, *Extremal problems for imbalanced edges*, Graphs Comb. **22** (2006) 103–111.